\documentstyle[epsf,psfig,11pt]{article}
\newcommand{\be}{\begin{eqnarray}}

\newcommand{\ee}{\end{eqnarray}}

\newcommand{\xp}{x_{\cal P}}

\title{
        \begin{flushright}
        {\normalsize BNL-NT-01/4\\}
        \end{flushright}
\bf  Deeply inelastic scattering off nuclei at RHIC}
\author{ Raju Venugopalan\\
        {\small\it Physics Department and RIKEN-BNL Research Center,
        BNL,
        Upton, NY 11973, USA } \\
         }
\date{}

\begin{document}

\maketitle

\begin{center}
{\bf Abstract}\\
\end{center}

We discuss the physics case for an electron--nucleus collider at 
RHIC.

\vfill \eject

\section{Introduction}

A high energy electron--nucleus collider, with a center of mass energy
$\sqrt{s}=60$--$100$ GeV, presents a remarkable opportunity to explore
fundamental and universal aspects of QCD. The nucleus, at these
energies, acts as an amplifier of the novel physics of high parton
densities--aspects of the theory that would otherwise only be explored
in an electron--proton collider with energies at least an order of
magnitude greater than that of HERA. An electon--nucleus collider will
also make the study of QCD in a nuclear environment, to an extent far
beyond that achieved previously, a quantitative science. 
In particular, it will help complement, clarify, and reinforce physics
learnt at high energy nucleus--nucleus and proton--nucleus collisions
at RHIC and LHC over the next decade. For both of these reasons, an eA
collider facility represents an important future direction in high
energy nuclear physics.

We will summarize here the physics arguments that support both the key
points above. We will also briefly discuss experimental observables in
deeply inelastic scattering (DIS) and signatures of novel
physics. Accelerator and detector issues have been discussed
elsewhere.  Details on these, on the physics issues, and references to
an extensive literature can be found in 
proceedings~\cite{proceed1,proceed2} of two of the
three eRHIC workshops that were held in the last year~\footnote{More
information on eRHIC and on previous eA studies for HERA can also be
found at the website: http://quark.phy.bnl.gov/ raju/eRHIC.html} and 
in the earlier proceedings of eA HERA workshops~\cite{Arneodoetal}.

The physics arguments can be separated according to the kinematic regions of 
interest~\footnote{$m_N$ below is the nucleon mass.}. Very roughly, 
these are

\begin{itemize}
\item
the small $x_{Bj}$ region ($x_{Bj} < 1/(2m_N R_A) \approx 0.01$ for a large 
nucleus), where 
the virtual photon interacts coherently with partons in a nucleus over
a region exceeding its longitudinal extent $2 R_A$.

\item
the intermediate $x_{Bj}$ region ($1/(2m_N R_A) < x_{Bj} < 1/(2m_N R_N)\approx 
0.1$ for a large nucleus), where the virtual photon interacts 
coherently over longitudinal 
distances larger than the longitudinal size of the nucleon $2 R_N$, 
but smaller than the longitudinal size of the nucleus $2 R_A$.

\item
the large $x_{Bj}$ region ($x_{Bj}> 1/(2m_N R_N)\approx 0.1$ for a large 
nucleus) where the 
virtual photon/W or Z boson is localized within a longitudinal distance 
smaller than the nucleon size.

\end{itemize}

In this talk, we will cover only the physics of the small $x_{Bj}$ region.
Due to space limitations, we will not cover the interesting physics 
at intermediate $x_{Bj}$ that can be studied with an eA collider. 
This covers the region in $x_{Bj}$ from where inter--nucleon forces become 
important to coherent effects involving several nucleons. 
A nice discussion of these issues (in the context of the HERA eA 
collider proposal) can be found in Ref.~\cite{Arneodoetal}. We will 
not discuss the physics of the large $x_{Bj}$ region either--this 
topic has been covered by other participants at this meeting~\cite{largex}.

\section{eA physics at small $x_{Bj}$: $x_{Bj} < 1/(2 m_N R_A)$}

\begin{figure}[here]
\epsfxsize=10.cm 
\hfil
\epsffile{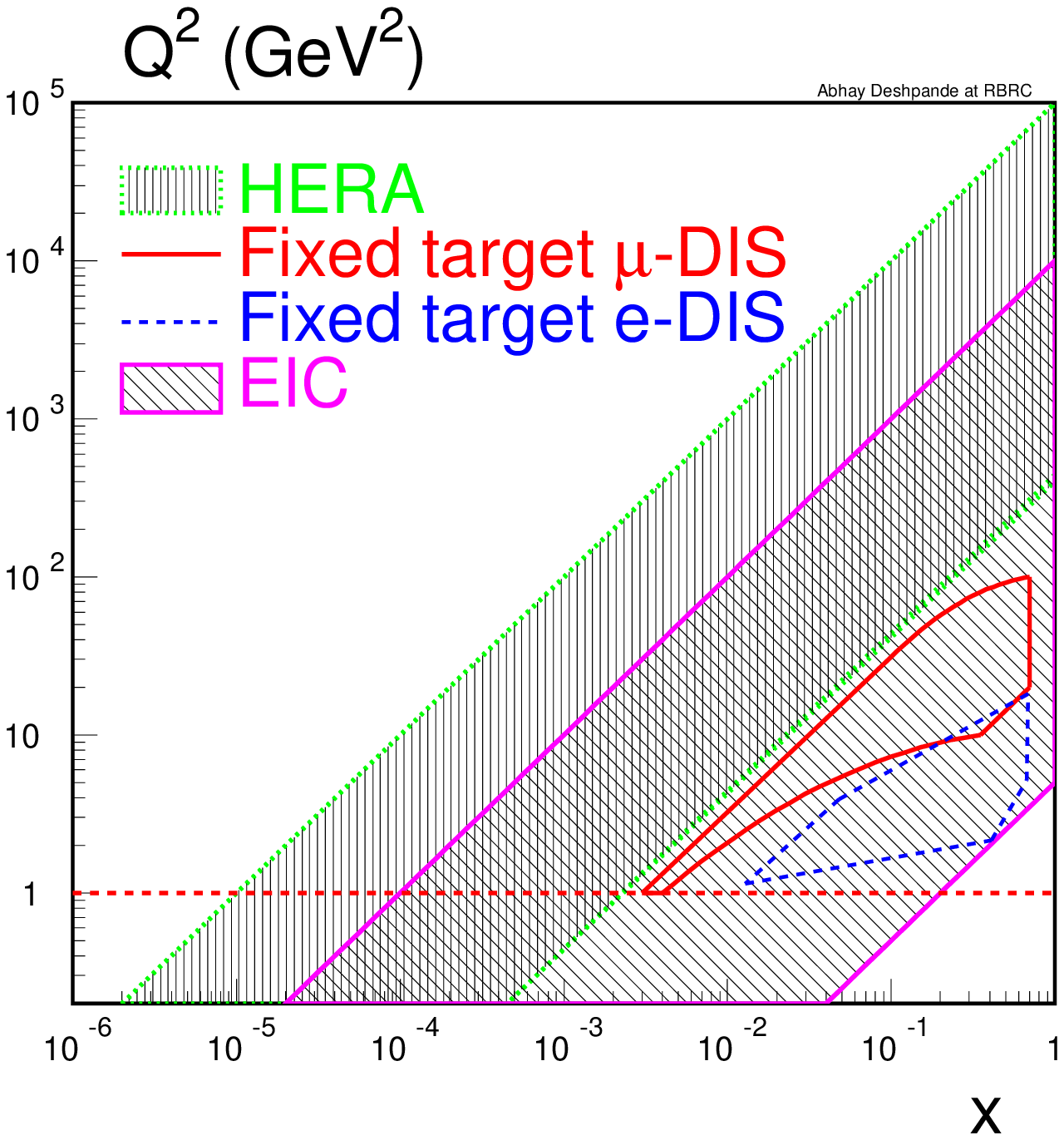}
\hfil
\caption{The $x$--$Q^2$ range of the electron ion collider (EIC) compared 
to that of the HERA ep collider and fixed target experiments. The EIC's 
reach would encompass the fixed target regime as well as part of the 
HERA regime.}
\label{fig:xq2}
\end{figure}

This regime of small $x_{Bj}$'s ($x_{Bj}\leq 0.01$) is easily accessed
by an electron--heavy ion collider in the energy range
$\sqrt{s}\approx 60$--$100$ GeV. Fig.~\ref{fig:xq2} 
is a plot of the $x-Q^2$ plane
delineating the range mapped for $\sqrt{s}=63$ GeV ($10$ GeV electrons
on $100$ GeV heavy ions at RHIC). What is novel about these energies
is that for the first time one can study the physics of $x_{Bj}<<
0.01$ in a nucleus for $Q^2 >> \Lambda_{QCD}^2$, where
$\Lambda_{QCD}\sim 200$ MeV.  Previous (fixed target) experiments such
as NMC and E665 and current ones such as HERMES and COMPASS 
could only access small $x_{Bj}$ at small $Q^2$'s.

Some questions that may come immediately to mind are: 

i) why is it
important to simultaneously have large $Q^2$ at small $x_{Bj}$ ?  
\vskip 0.05in
ii)
Hasn't HERA explored this $x_{Bj}$--$Q^2$ range already? What then 
can one learn by studying the same regime with an eA collider?
\vskip 0.05in
iii) In the
nuclear context, havn't the fixed target experiments at CERN, DESY and
Fermilab, studied the small $x_{Bj}$ regime? 
\vskip 0.05in
iv) Does the collider environment have a compelling 
advantage in the study of small $x_{Bj}$ physics?

In the following, we will address in detail the physics issues that
underly these queries. The pithy answer to all of these however is that 
{\it an eA collider in the desired energy range may probe a hitherto 
inaccessible regime
of QCD, where the properties of strongly interacting matter are
radically different from those studied previously}. Understanding the 
properties of QCD in this regime may provide us the answer to 
fundamental questions about the strong interactions that remain 
unanswered. A brief list of these open questions is:
 a) what is the nature of multi--particle production? b) how do 
cross--sections behave at high energies? Are the bulk features of 
the cross--section computable in QCD? c) Are the properties of 
hadrons universal at very high energies? d) what is the nature of 
confinement-in particular, as probed in striking phenomena such 
as hard diffraction? and e) what are the initial conditions for 
heavy ion collisions, and how do they affect the formation of a 
quark gluon plasma?

We will also
emphasize that mapping the relevant $x-Q^2$ regime with the proposed collider,
at the high luminosities considered, will provide measurements of
several physical quantities, with a much higher degree of precision,
and of course in a wider kinematic range. Aside from their intrinsic
interest, these quantities will be extremely important for the physics
goals of other current and future experiments.

\subsection{Why is it important to simultaneously have large $Q^2$ and 
small $x_{Bj}$?}

In deeply inelastic scattering (DIS), one has the exact kinematic relation 
\be
y\, x_{Bj}= Q^2/s \, .\nonumber 
\ee
All of these variables are invariants--they are
frame independent. The invariants $x_{Bj}$ and $Q^2$ are of course well known 
-- they are simply related, respectively, to the fraction of the momentum of 
a hadron or nucleus carried by a parton, and to the momentum transfer 
squared from the electron to the hadron. The invariant $y$, in the 
rest frame of the target, is the ratio of the energy transferred to 
the hadron to the energy of the electron. It has the kinematic 
range $0\leq y \leq 1$. For the purposes of this discussion, we 
will assume that $y\sim 1$, or $x_{Bj}\sim Q^2/s$~\footnote{How large 
a value of $y$ can be obtained without being swamped by uncertainities 
in the radiative corrections is a very important technical 
issue we will not address 
here. It has been addressed previously in proceedings of the eA at HERA 
workshops. These can be accessed on the World Wide Web at the URL: 
http://www.desy.de/~heraws96/proceedings/}.

The physics of small $x$ is the physics of high energies~\footnote{For a 
review of recent theoretical developments, see Ref.~\cite{Pramana}.}. 
The total cross--section in strong interaction physics can be parametrized 
by the power law behavior $\sigma (s) \sim s^\epsilon$,  where 
$\epsilon\sim 0.08$. Thus the cross--section grows with decreasing $x$ 
and, at high energies, is dominated by small $x$. This behaviour is 
explained in Regge phenomenology via Pomeron exchange--the t--channel 
exchange of an object with vacuum quantum numbers. Though the Pomeron 
hypothesis has had some striking success~\cite{DL} 
in explaining high energy data, 
and has been the paradigm for understanding non--perturbative 
multi--particle production, 
it is not clear that it can be interpreted as an actual particle and 
understood as arising from the fundamental theory. A popular 
construction, first postulated by Francis Low and Shmuel Nussinov~\cite{LowNussinov}, is that the 
Pomeron is two gluon exchange with vacuum quantum numbers in the $t$  channel. However, since 
total cross-sections at lower energies than available currently 
were dominated by very soft transverse momenta, it proved very hard to 
come up with a robust QCD based theory of the Pomeron (or more generally,  
that of the behavior of the bulk of the cross-section at high energies), 
that would also have predictive power.

The situation has changed with the advent of colliders at very high
energies. With the Hadron Electron Ring Accelerator (HERA) at DESY, 
where $27.5$ GeV electrons collide with $920$ GeV
protons, corresponding to a center of mass energy $\sqrt{s}\sim 300$
GeV, one can have $Q^2 = 1$--$10$ GeV$^2$ for $x_{Bj}\sim 10^{-4}$. In
prior experiments, at these low $x_{Bj}\sim 10^{-4}$, only $Q^2\ll
\Lambda_{QCD}^2$ could be accessed. Thus even though one was probing
the small $x_{Bj}$ regime of the Pomeron, the coupling constant was
too large to make predictions and therefore test/extract information
about the theory in this regime. Since QCD is enormously complex, the 
lack of a small parameter in this regime was problematic. 
It  hobbled progress in small $x_{Bj}$ physics even 
though a wealth of tantalizing small $x_{Bj}$ data~\cite{Arneodoetal} exists at 
small $Q^2$. A large number of models were constructed to 
understand the data, but their connection to the fundamental theory is 
still tenuous.

At HERA, the coupling $\alpha_S(Q^2)\ll 1$ in a significant portion of the 
small $x_{Bj}$ regime of interest. Since the coupling is weak, computations 
can be made in pQCD and tested against the data. 
It lead, for instance, to the resurrection of the idea of the perturbative 
Pomeron developed by Lipatov and colleagues in the late 1970's-now known 
by the acronym BFKL Pomeron~\cite{Lipatov}. In QCD, 
a Pomeron can be constructed from the exchange of gluon ladders--the 
so--called hard Pomeron. The leading order BFKL result 
predicts rising cross--sections that rise more rapidly than the HERA data 
support. The next to leading order correction is very large and negative, 
thereby causing great confusion (and interest) in the QCD 
community~\cite{FadLip,CiaCam}. 
One possibility is a more subtle resummation of next-to-leading order 
small $x$ effects in the BFKL framework~\cite{Salam}; another is to formulate the 
problem of QCD at small $x_{Bj}$ in the language of high parton densities--thereby performing a different sort of resummation~\cite{GLR,MQ,MV}.
This topic will be addressed in the following sub--section.

The ability to probe large $Q^2$'s at small $x_{Bj}$ has thus given us 
a handle on understanding, in a quantitative way, the 
hitherto inaccessible 
small $x_{Bj}$ regime of QCD. This represents tremendous progress since it 
is this regime of the theory that controls the bulk of high energy 
cross--sections. Without understanding this regime, one cannot claim 
reliably that one completely understands the theory.

In what follows, we will discuss what we have learnt from 
the HERA experiments in the small $x_{Bj}$ and large $Q^2$ regime, and 
how these experiments point to novel physics that may be fully explored 
with an eA collider.

\subsection{From HERA towards a new regime of high parton densities}

\begin{figure}
\epsfxsize=9.0cm
\epsffile{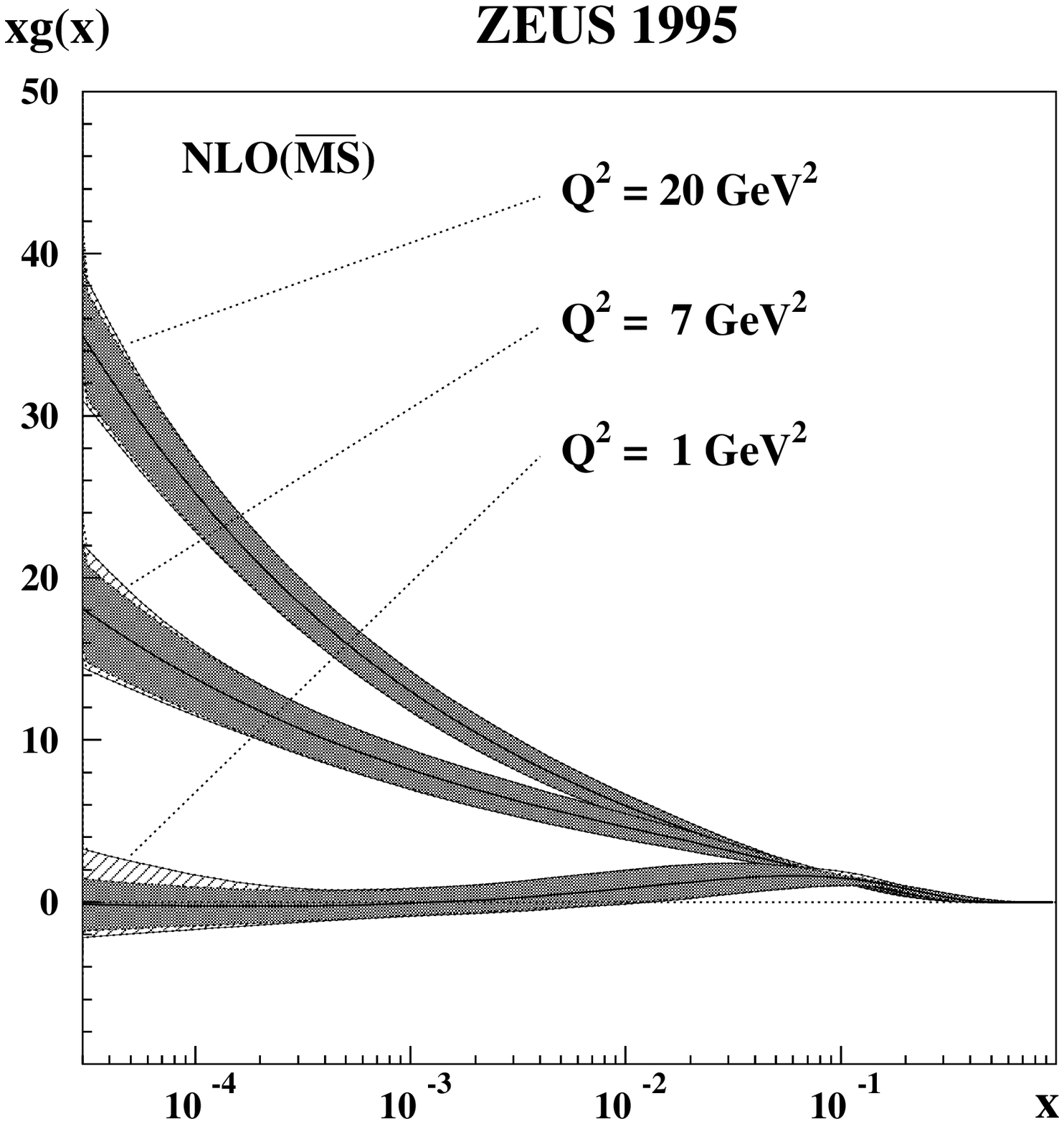}
\caption{The gluon distribution for three different values of $Q^2$ extracted 
using the ZEUS NLO QCD analysis. From Ref.~\cite{ZEUSa}.}
\label{fig:3Q2}
\end{figure}

The wide kinematic range in $x$ and $Q^2$ 
of the HERA collider can be seen in Fig.~\ref{fig:xq2}. 
One of the striking results 
from HERA is that the gluon distribution, extracted from scaling 
violations of $F_2$, grows rapidly at small $x$ and 
high $Q^2\gg \Lambda_{QCD}^2$. This is shown in Fig.~\ref{fig:3Q2}.
 This tells us 
that at high energies the proton is not a simple object with three valence 
quarks and a few gluons that bind together the quarks. 
At a fixed external scale $Q^2=20$ GeV$^2$, one finds $25$--$30$  gluons, 
per unit rapidity, at $x_{Bj} = 10^{-4}$ 
in the proton. The proton is therefore very rapidly 
growing dense as the resolution scale in $x$ is shifted to  smaller $x$'s. 

At high $Q^2$, ($Q^2\gg 10$ GeV$^2$) the rise in the gluon structure
function at small $x$ is very well understood~\cite{DeRujula} in the
framework of perturbative QCD (pQCD). An asymptotic expression for the
rise is the double logarithmic formula where the gluon distribution
grows as
\be
G(x,Q^2)\sim 
\exp\left[\sqrt{\ln\left(\ln\left({Q^2\over \Lambda_{QCD}^2}\right)\right)\ln(1/x)}\,\right] \, ,
\ee
This double logarithmic behavior was tested at HERA. It is claimed
that the value of $\alpha_S(Q^2)$ extracted from the fit provides a
precise determination of the coupling at $M_Z\sim 91$ GeV and is in
agreement with other world data~\cite{BallForte}.  More detailed NLO
QCD fits with different sets of parton distributions~\cite{PDF} have
been shown to describe the HERA data for a wide range of $x_{Bj}$ and
$Q^2$.  At high $Q^2$, the deeply inelastic scattering data is a
testament to the striking success of perturbative QCD.

In the very high $Q^2$ regime--$Q^2\gg 10$ GeV$^2$, one is not probing
the region of extremely small $x_{Bj}$: for $Q^2=10$ GeV$^2$, the
smallest $x_{Bj}$ available is $\sim 10^{-4}$. In the region of $Q^2=
1$--$10$ GeV$^2$, at correspondingly smaller $x_{Bj}$, the situation
from the usual pQCD standpoint is less
clear~\cite{ManSaaWei,AltBalFort}. The HERA ZEUS and H1 QCD fits agree
with the data but with the price being that one extracts an
anomalously small value of the gluon distribution, and one obtains
more sea quarks than glue at small $x$~\cite{ZEUSa,H1a}. The
anomalously small gluon distribution is seen in Fig.~\ref{fig:3Q2} for
$Q^2=1$ GeV$^2$ where, at small $x$, the distribution is consistent
with zero.  Several groups have argued that one obtains results that
run contrary to our intuition because the standard pQCD approach is
breaking down~\footnote{For a summary of recent discussions, see
Ref.~\cite{McDermott}.}.  The Tel Aviv group of Gotsman et al., for
instance, claims that there is no pQCD fit that can simultaneously
explain the inclusive $F_2$ data and the large amount of data on the
energy dependence of $J/\psi$ photo--production~\cite{TelAviv}. For a
recent discussion of unitarity and long distance effects in $J/\psi$
photo--production, see Ref.~\cite{FDS}.

The argument is that screening effects due to large parton densities 
are important in this regime and have to be taken into account. The 
physics is still weak coupling though; one still has 
$\alpha_S \ll 1$ in the $Q^2=1$-$10$ GeV$^2$ regime. Phenomenological 
models 
that take these effects into account, and match into the usual pQCD 
formalism at high $Q^2$,  have been successful in fitting both the 
inclusive and the diffractive HERA data~\cite{GBW,GLM,FGS,BKS}.

There are therefore tantalizing hints from the HERA data that one is 
beginning to see the effects of large parton densities in the proton.
We will argue below that standard pQCD breaks down when the parton 
densities become very large. Even though the coupling is 
weak, the physics will be non--perturbative due to the high field 
strengths generated by the large number of partons. This is a 
novel regime of the theory. We will further argue that an eA 
collider is much better suited to explore this regime even though its
$x$-$Q^2$ range will be somewhat less extensive than that achieved at 
HERA.

\subsection{QCD is a colored glass condensate at high energies}

In the infinite momentum frame (IMF), 
the number of partons per unit transverse 
area, for a fixed resolution of the external probe $Q^2\gg \Lambda_{QCD}^2$, 
grows rapidly with the energy--or with decreasing $x_{Bj}$. In this 
high parton density regime, the corresponding 
QCD field strength squared~\footnote{$F_{\mu\nu}^2$ is frame independent.}  
becomes $F_{\mu\nu}^2\sim 1/\alpha_S$: since $\alpha_S(Q^2)\ll 1$, the 
color field strengths in this regime are large~\cite{Mueller1}. The non--linearities 
inherent in the theory become manifest, radically altering the properties 
of distributions in high energy collisions. For instance, the gluon 
distribution that was growing slowly now saturates--and grows very slowly--
at most logarithmically with decreasing $x_{Bj}$.

In this high parton density--large field strengths-- regime, the 
saturated gluons,
when viewed in the IMF, form a novel state of matter which we will
henceforth call a color glass condensate (CGC)~\cite{Larry}. Why a glass, and why a
condensate? At small $x_{Bj}$, most of the partons are rapidly
fluctuating gluons that interact weakly with each other. They are
however strongly coupled to the large $x_{Bj}$ ``hard'' parton color
charges, that act as random, static, sources of color charge. This is
exactly analogous to a glassy system--in particular, one can show that
there is a formal analogy to spin glass condensed matter systems~\cite{ParisiSourlas}.  
In the latter case, one
has a disordered state of spins coupled, say, to random magnetic
impurities --in the ``quenched'' limit, these impurities are
static--or long lived.

Further, since the occupation number of the gluons is large, they form 
a condensate. Being bosons, arbitrary numbers of gluons can pile up in 
a momentum state. In the classical Bose gas, for instance, Bose--Einstein 
condensation leads to a dramatic overpopulation of the zero momentum 
state. In our case, since the gluons are interacting, and have both 
attractive and repulsive interactions, they ``pile'' up in a narrow 
band of states, peaked at a typical momentum we shall call the 
``saturation'' momentum~\cite{MV,JKMW,Kovchegov,KovMuell}. The saturation scale at a particular $x$ is 
the density per unit area of all the parton sources at higher $x$'s. 
One has
\be
Q_s^2 (x) = \frac{1}{\pi R^2}\, \frac{dN}{d\eta} \, ,
\ee
Here $\eta=\ln(1/x)$ is the rapidity. As the energy increases, or
$x_{Bj}$ decreases, this ``bulk'' scale of the condensate
grows and one can have $Q_s \gg \Lambda_{QCD}$. The distinction between fields and sources is of
course arbitrary-as one decreases $x$, what were formerly fields turn
into sources--thereby increasing the density of sources. This
transformation is nothing but the ``block spin'' renormalization group
transformation of Wilson, and the equations describing the evolution
to small $x_{Bj}$ are Wilsonian renormalizaton 
group equations~\cite{JKMW,JKW}. There has been significant theoretical 
progress recently in understanding the asymptotic behavior of these 
renormalization group equations~\cite{RG}.

Thus, because a large scale $Q_s$ is generated at small $x$, 
one can  predict the
behavior of this non--perturbative condensate using weak coupling QCD
techniques.  An interesting question we don't have the answer to 
yet is how the coupling constant behaves in this regime--is 
there a fixed point of the theory at high energies?
It would be therefore be absolutely remarkable, and of fundamental interest, if it could be 
demonstrated empirically that QCD at very high energies is a non--trivial glassy condensate of gluons.

\begin{figure}
\epsfxsize=9.0cm
\epsffile{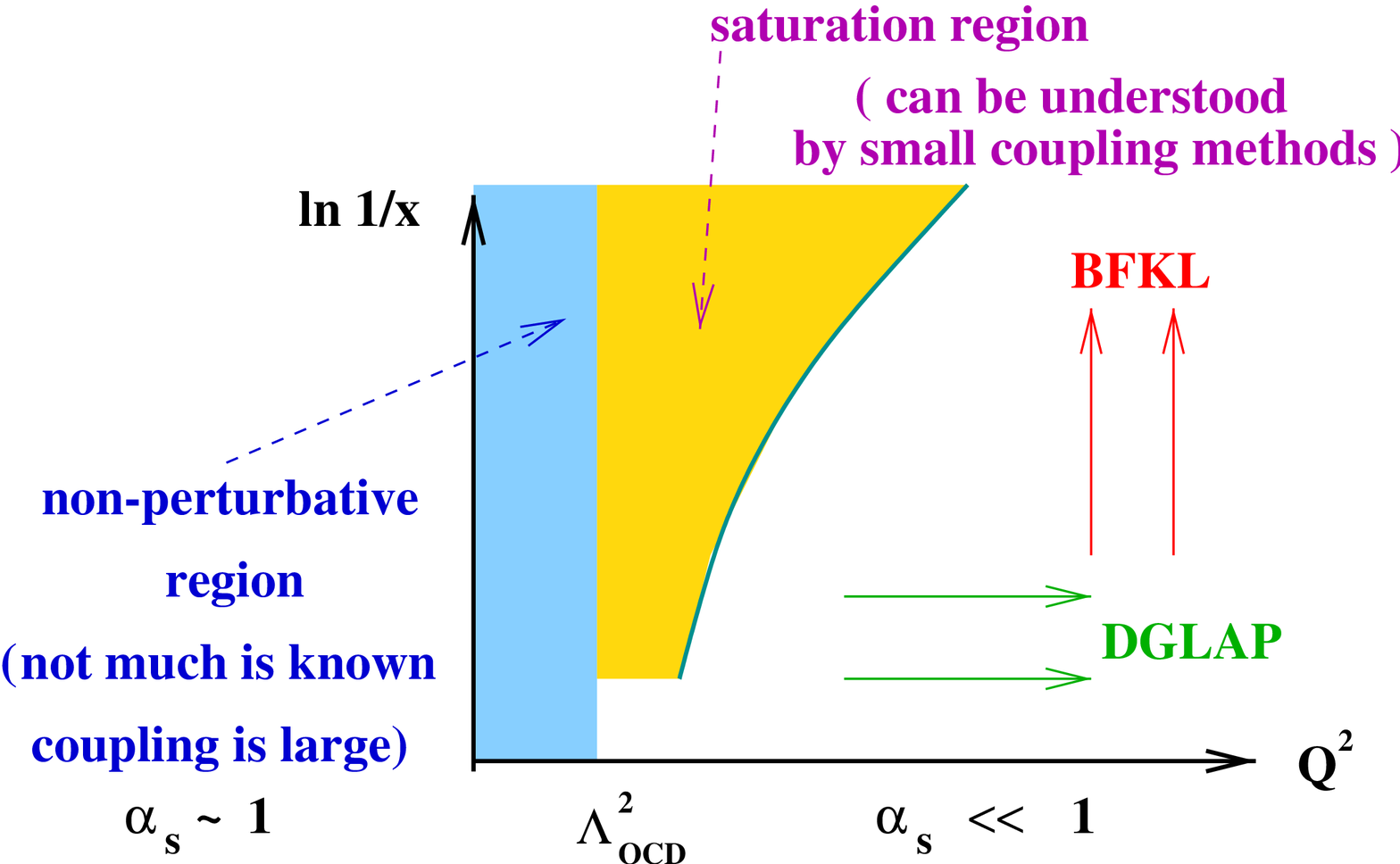}
\caption{Schematic diagram of the $\ln(1/x)$--$Q^2$ plane conveying a 
rough idea of the different regimes of applicability of the different 
evolution equations. Figure courtesy of Y. Kovchegov.}
\label{fig:yura}
\end{figure}

In Fig.~\ref{fig:yura}, is plotted a (very) schematic diagram of scattering
in the $\eta=\ln(1/x)$ versus $Q^2$ plane. If $x_{Bj}$ is not too
small, and $Q^2$ is large, the
Dokshitzer--Gribov--Lipatov--Altarelli--Parisi (DGLAP) QCD evolution
equations~\cite{DGLAP} work very well. For a fixed $Q^2$, the
Balitsky--Fadin--Kuraev--Lipatov (BFKL) equations describe the $x$
evolution of distributions in a limited kinematic range. Both of these
are linear evolution equations and do not fully take into account the
non--linearities of the theory. Indeed, with regard to the latter, it
is not clear there is a physical kinematical region available for
linear evolution, where these equations apply, before high parton
density effects set in.  The line in the $\eta$-$Q^2$ plane represents
the scale $Q_s(x)$ that separates the extensively studied regime of the 
well known QCD evolution equations from the saturation regime of the CGC.

We discussed in the previous sub--section how the HERA data may be showing 
hints of screening effects that may be the precursor to the saturation 
regime. We will argue below that deep inelastic scattering off large nuclei 
at high energies may be sufficient to probe this novel regime of the 
theory.

\subsection{Probing the colored glass condensate in eA DIS}

At a high energy eA collider, with energies $\sqrt{s}=60$--$100$ GeV,
one will access (roughly) $x_{Bj}=10^{-4}$--$10^{-3}$ for, respectively,
$Q^2=1$--$10$ GeV$^2$--see Fig.~\ref{fig:xq2}. 
These values of $x_{Bj}$ and $Q^2$
are in the ballpark (even if more limited in range) than those at
HERA. However, an eA collider has a tremendous advantage--the parton
density in a nucleus, as experienced by a probe at a fixed energy, is
much higher than what it would experience in a proton at the same
energy. Since the parton density grows as $A^{1/3}$, this effect is more
pronounced for the largest nuclei.  To probe a comparable parton
density in a nucleon, the probe would have to be at much higher
energies than presently available.

The physics behind this effect is subtle and is a result of quantum 
coherence. In DIS at small $x_{Bj}$, in the target rest frame, 
the virtual photon splits into a quark--anti-quark pair, that subsequently 
interacts with the nucleus. If $x_{Bj}\ll 1/(2 m_N R_A)\sim 0.01$, the 
$q\bar{q}$ pair interacts coherently with partons along the entire 
length of the nucleus. Furthermore, equally importantly, if the 
transverse separation of the pair $\sim 1/Q$ is smaller than the 
nucleon size ($Q^2 > \Lambda_{QCD}^2$), the probe will experience, coherently,
random $p_t$ kicks from partons in different nucleons along its trajectory. 
While $\langle p_t \rangle \sim 0$, fluctuations will be large: 
$\langle p_t^2\rangle \sim A^{1/3}$. In the IMF, this effect is 
interpreted as the $q\bar{q}$--pair experiencing large fluctuations of 
color charge in the nuclear ``pancake''. 
It is clear, from both viewpoints, that a large scale, proportional to 
the parton density per unit area, is generated in large nuclei due to 
quantum mechanical coherence at small $x_{Bj}$ and large $Q^2$. This 
scale is none other than the saturation scale $Q_s(x)$ discussed 
previously.

In a nucleus, one defines  
\be
Q_s^2 = {1\over \pi R^2}\, {dN\over dy} \equiv {A^{1/3}\over x^{\delta}} 
\, {\rm{fm}}^{-2} \, ,
\label{SAT}
\ee
Here $\delta$ is the power of the rise in the gluon distribution 
in a {\it nucleon} at the typical $Q^2\sim Q_s^2$ of interest~\footnote{
This scale must be determined self-consistently.}. At HERA, for 
$Q^2$ of a few GeV$^2$, a reasonable estimate is $\delta\sim 0.3$. Now if 
we ask at what $x_{Bj}$ in a proton  will the probe see the same parton 
density as in a nucleus, Eq.~\ref{SAT} suggests,
\be
x_{\rm {proton}} = {x_{\rm {nucleus}}\over 
\left( A^{1\over 3}\right)^{1/\delta}} \, .
\ee
Since the nucleus is dilute, and if, being conservative, we assume that 
one can't tag on impact parameter-we take the effective 
$A^{1/3}=4$, then for $\delta\sim 0.3$, we find $x_{\rm{proton}} \sim 
x_{\rm{nucleus}}/100$. Thus, one would obtain the same parton density in 
a nucleus at $x_{Bj}\sim 10^{-4}$ and $Q^2\sim$ a few GeV$^2$, as 
 would be attained in a {\it nucleon} at $x_{Bj}\sim 10^{-6}$ and similar 
$Q^2$! Put differently, it would take an electron--proton collider 
with an order of magnitude larger energy than HERA to achieve the 
same parton density as would be achieved by eRHIC. It is now believed 
that impact parameter tagging is feasible by counting knock-out 
neutrons~\cite{StrikZhalov,Krasny}--if so, the gain in parton density in eA relative to ep 
would be even more spectacular.

The small $x_{Bj}$ regime has been studied previously in fixed target
DIS off nuclei at CERN and Fermilab. In these experiments, the center
of mass energy was a factor of $3$--$5$ less than the proposed
collider. This corresponds to a factor of $10$--$25$ smaller in $Q^2$
for the same $x_{Bj}$.  It was therefore difficult to interpret the
experimental results at small $x_{Bj}$ in the framework of
perturbative QCD. Remarkable phenomena, such as shadowing, were
observed-the relation of the experimentally measured shadowing to the
physics of parton saturation presented here is at present unclear and
deserves to be explored further.  It could not be explored at the
fixed target experiments because the kinematics corresponded to an
intrinsically non--perturbative regime of the theory that is not
amenable to a weak coupling perturbative QCD based analysis upon
which the parton saturation picture rests.

In the following, we will discuss both inclusive and semi--inclusive 
signatures of the CGC. The latter, in particular, are striking. In this 
regard as well, the collider environment holds a significant edge since 
semi-inclusive observables proved very difficult to measure in 
fixed target eA DIS.

\subsection{Signatures of the new physics of the CGC}

A  number of inclusive and semi--inclusive experimental observables 
exist that will be sensitive to the new physics in the regime of high 
parton densities. All the inclusive and semi--inclusive 
observables that were studied at HERA can be studied with eRHIC--with 
a ZEUS/H1 type detector design~\cite{Krasny}. 
However, due to the remarkable versatility 
of RHIC, and due to likely improvements in detector design, several 
new observables can be measured in the small $x_{Bj}$ region for the 
first time. We will first discuss inclusive variables and the 
signatures of new physics in these. We will then discuss semi-inclusive 
observables.

\vskip 0.05in

\noindent\underline{Inclusive signatures of the CGC}
\vskip 0.05in

An obvious inclusive observable is the structure function
$F_2(x_{Bj},Q^2)$ and its logarithmic derivatives with respect to
$x_{Bj}$ and $Q^2$.  eRHIC should have sufficient statistical
precision for one to extract the logarithmic derivatives of
$F_2$ (and of its logarithmic derivatives!). Whether the systematic errors at small $x_{Bj}$ will affect the
results is not clear at the moment.

\begin{figure}[here]
\epsfxsize=10.cm 
\hfil
\epsffile{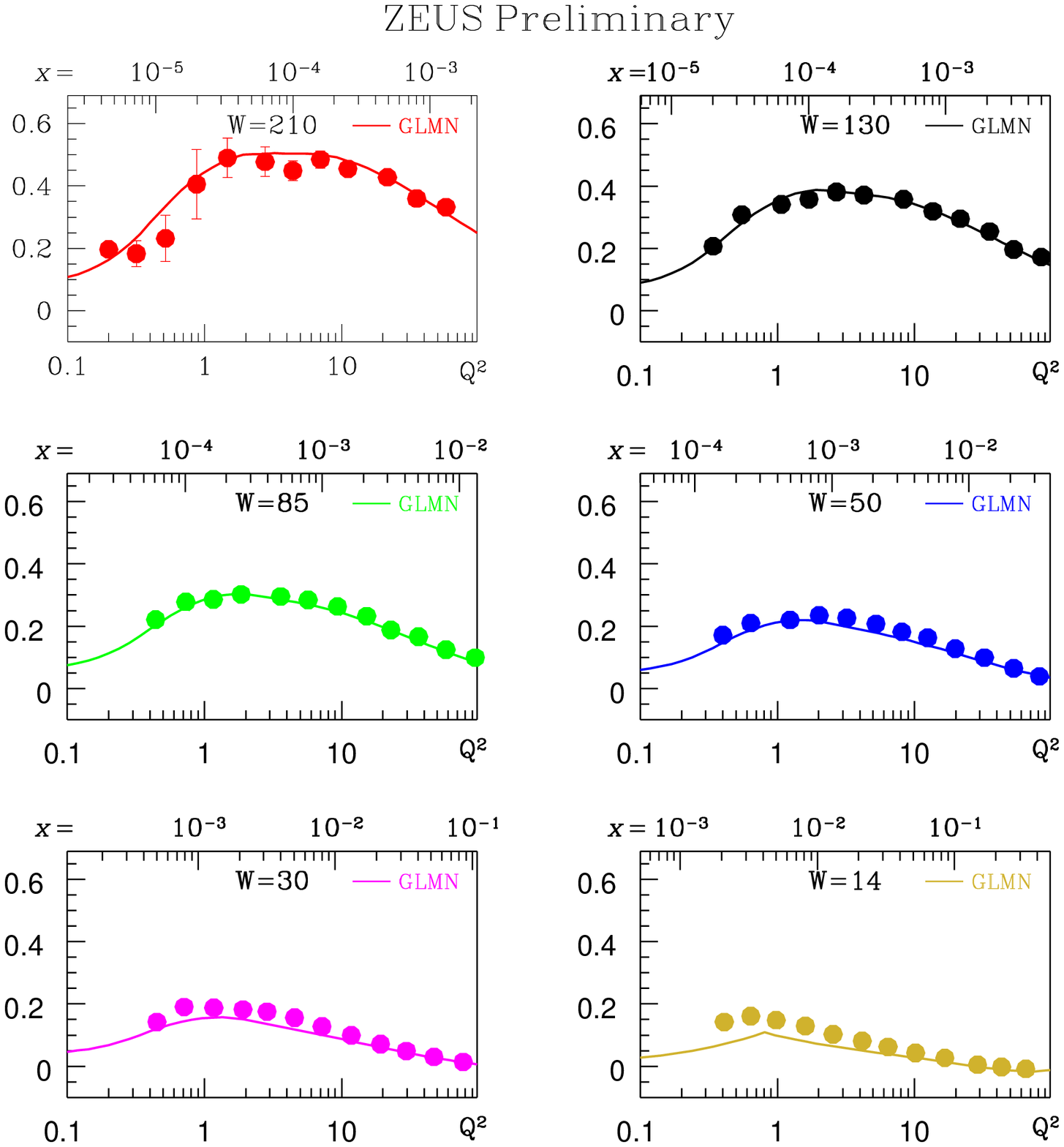}
\hfil
\caption{The slope $dF_2/d\ln(Q^2)$ versus $Q^2$ for fixed $W^2$. Figure 
from Ref.~\cite{TelAviv}.}
\label{fig:slopeW}
\end{figure}

\begin{figure}[here]
\epsfxsize=10.cm 
\hfil
\epsffile{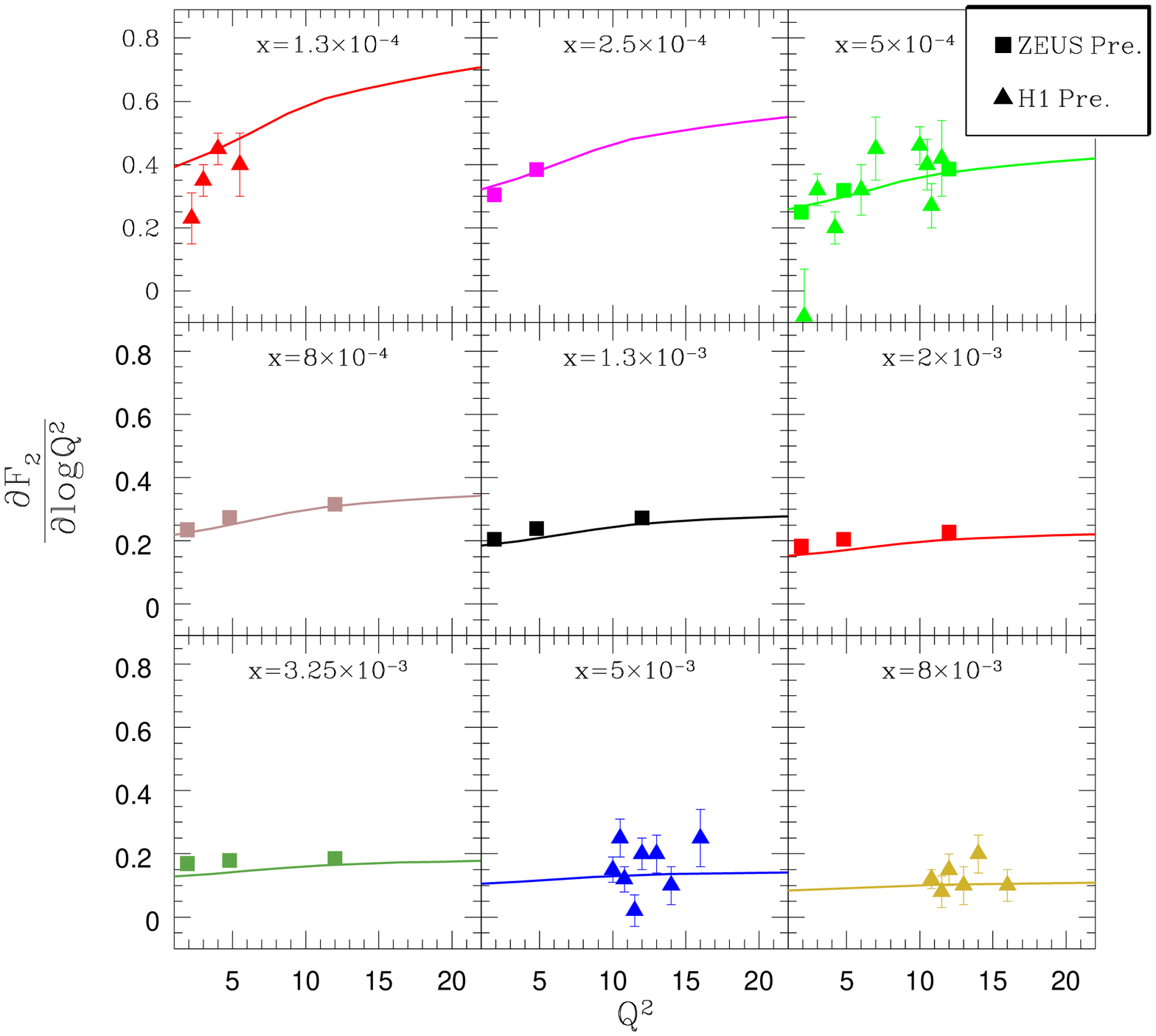}
\hfil
\caption{The slope $dF_2/d\ln(Q^2)$ versus $Q^2$ for fixed $x$. }
\label{fig:slopex}
\end{figure}

The logarithmic derivative $dF_2/d\ln(Q^2)$, at fixed $x_{Bj}$, and
large $Q^2$, as a function of $Q^2$, is the gluon distribution. QCD
fits implementing the DGLAP evolution equations should describe its
behavior at large $Q^2$.  At smaller $Q^2$, one should see a
significant deviation from linear QCD fits--in principle, if the $Q^2$
range is wide enough, one should see a turnover in the
distribution. The $Q^2$ at which the turnover takes place should be
systematically larger for smaller $x$'s and for larger
nuclei. Predictions for this quantity, as a function of $Q^2$, for
fixed $W^2$ and for fixed $x$, in a phenomenological model, are shown
in Figs.~\ref{fig:slopeW} and \ref{fig:slopex} respectively.

A remarkable feature of eRHIC will be that one can extract the longitudinal 
structure function $F_L(x_{Bj},Q^2)= F_2 - 2\, x_{Bj} F_1$ at small $x_{Bj}$  
for the first time. An independent extraction of $F_L$ requires 
that the energy of the colliding beams be varied significantly. At RHIC,  
this is feasible. In the parton model, $F_L=0$--thus $F_L$ is 
very sensitive to scaling violations. In particular, it provides an 
independent measure of the gluon distribution-a fact that makes this 
quantity very important to measure in its own right~\cite{Wilczeketal}.

\begin{figure}
\epsfxsize=9.0cm
\epsffile{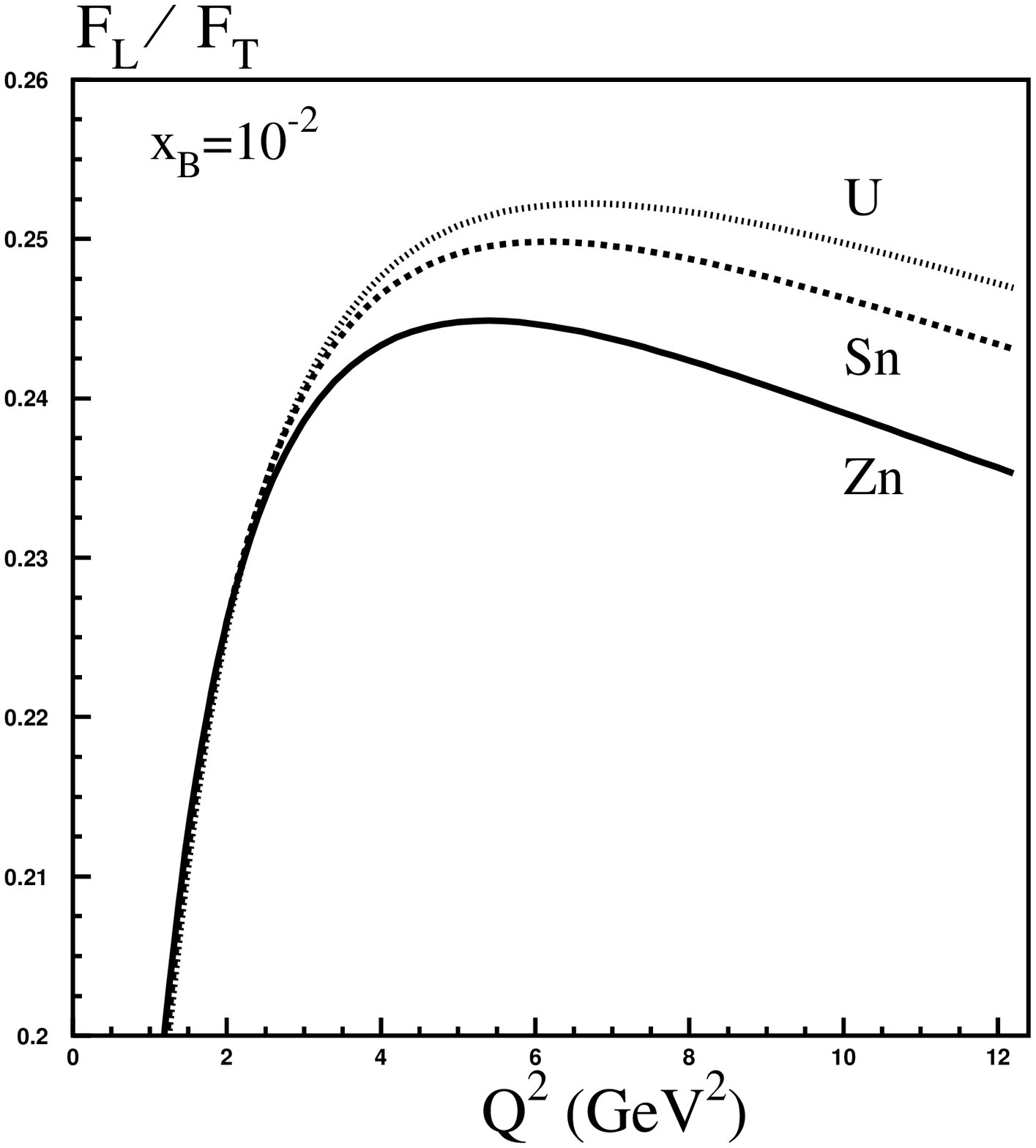}
\caption{The ratio $F_L/F_T$ as predicted in Ref.~\cite{BNLTA}. This ratio 
is plotted as a function of $Q^2$ for different nuclei and for fixed 
$x_{Bj}$.}
\label{fig:lt2}
\end{figure}

\begin{figure}
\epsfxsize=9.0cm
\epsffile{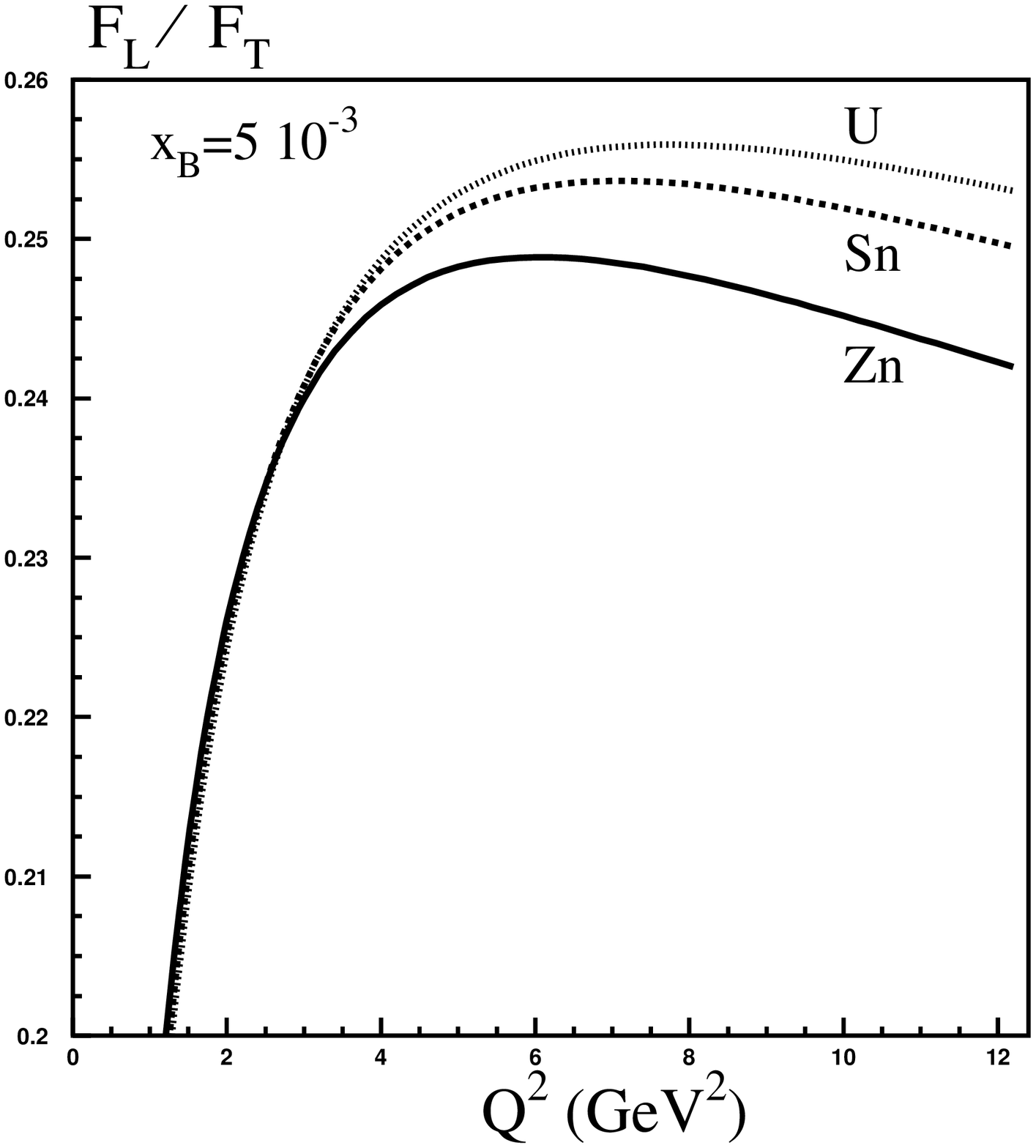}
\caption{The ratio $F_L/F_T$ as predicted in Ref.~\cite{BNLTA}. This ratio 
is plotted as a function of $Q^2$ for different nuclei and for a different 
$x_{Bj}$ than Fig.~\ref{fig:lt2}.}
\label{fig:lt3i}
\end{figure}

It has been suggested by several authors that $F_2 = F_L + F_T$ is not
very sensitive to higher twist saturation effects which may be
prominent in both $F_L$ and $F_T$ but may cancel in the
sum~\cite{Bartels}.  An independent measurement of $F_L$, and thereby
of $F_T$, will confirm this claim. The ratio of $F_L/F_T$ has a very
particular behavior in screening/saturation models. In
Figs.~\ref{fig:lt2} and \ref{fig:lt3i} is shown the prediction from a
particular model for this ratio~\cite{BNLTA}. The ratio
$F_L/F_T$, for a fixed $x_{Bj}$, has a maximum at a particular $Q^2$;
this maximum grows with the nuclear size (Fig.~\ref{fig:lt2}). As
$x_{Bj}$ decreases, the position of the maximum, for each nucleus,
increases (Fig.~\ref{fig:lt3i}). The maximum at which the turnover
$Q_{eff}(x,A)$ occurs is related to the saturation scale $Q_s$. The precise
relation is not known currently independently of particular models. To 
study the $A$ dependence, it might be more useful to look at $F_L$ and 
$F_T$ separately.

A very important inclusive observable is nuclear shadowing. 
Quark shadowing is defined through the measured ratio of the nuclear 
structure function $F_2^A$ to $A$ times the nucleon structure function 
$F_2^N$: $S_{\rm{quark}}=F_2^A/A F_2^N$. Gluon shadowing is similarly 
defined to be $S_{\rm{gluon}} = G_A/ AG_N$. Quark shadowing was observed
in the fixed target experiments (NMC,E665$\cdots$) and gluon shadowing, 
indirectly, through logarithmic derivatives of $F_2$. However, the 
gluon shadowing data at the smallest $x$'s are also at very low $Q^2$--
where the application of perturbative QCD is unreliable.

\begin{figure}
\epsfxsize=9.0cm
\epsffile{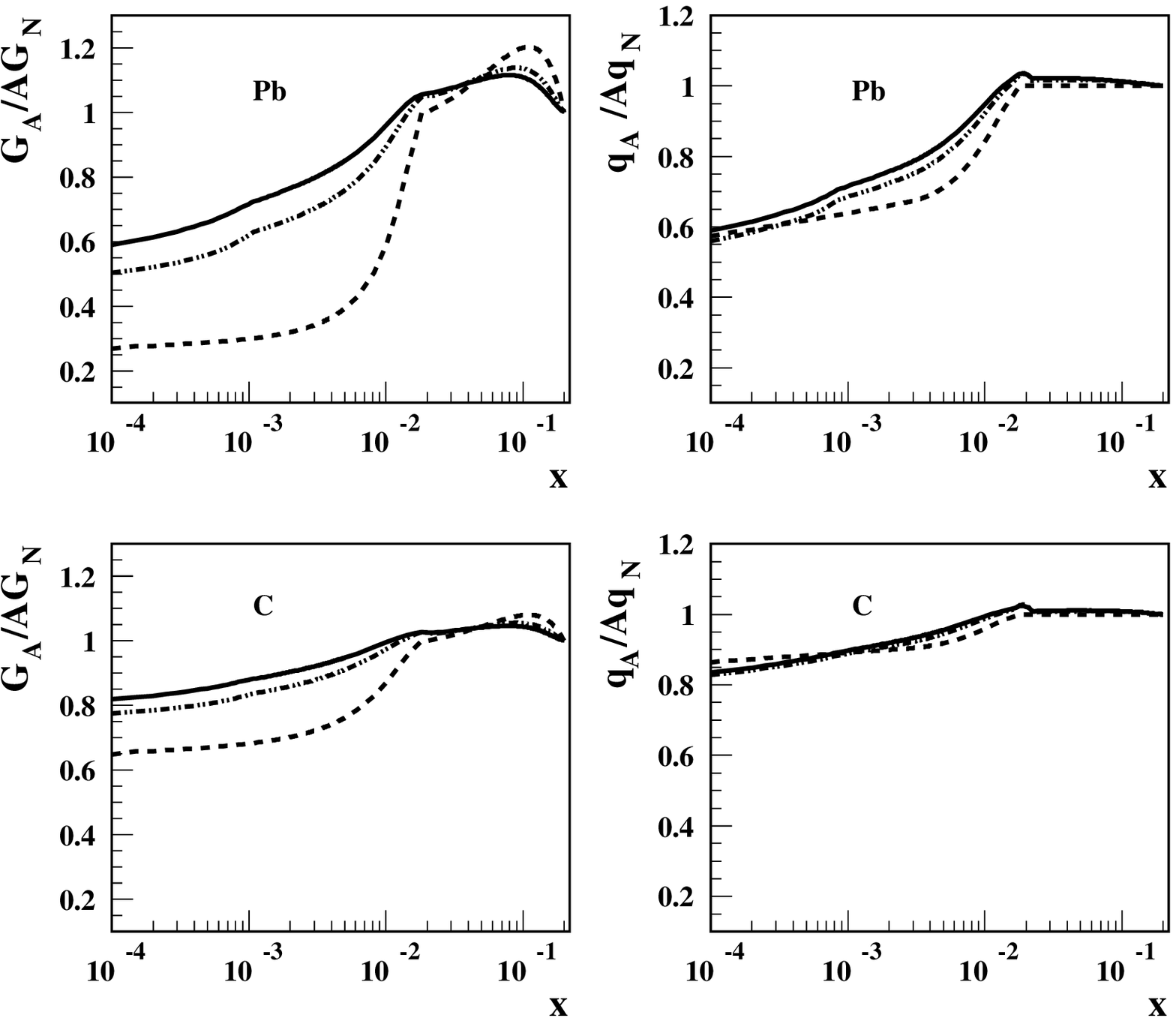}
\caption{Gluon shadowing $G_A(x,Q^2)/A G_N(x,Q^2)$ and quark 
shadowing $q_A/A q_N$ versus $x_{Bj}$ for lead (Pb) and Carbon (C). 
The different curves correspond to $Q = 2$ GeV (dotted), $Q=5$ GeV (dashed) 
and $Q=10$ GeV (solid). Figure from Ref.~\cite{FGS}.}
\label{fig:rqg}
\end{figure}

There are model calculations (see Fig.~\ref{fig:rqg}) that suggest
that gluon shadowing is very large at small $x_{Bj}$'s and fairly
large $Q^2$~\cite{FGS}.  eRHIC can help confirm if this is the
case. In addition, because of the extended kinematic range of eRHIC,
we can determine whether shadowing is entirely a leading twist
phenomenon, or if there are large higher twist perturbative corrections.  Some
saturation models, for instance, predict that perturbative shadowing
will become large as one goes to smaller $x_{Bj}$'s~\cite{JamalWang}. 
Isolating
perturbative contributions to shadowing from non--perturbative ones
will be an interesting experimental and theoretical challenge. Another
interesting question is whether shadowing saturates at a particular
value of $x_{Bj}$, for fixed $Q^2$ and $A$. How does this value vary
with $x_{Bj}$ and $Q^2$? Does the ratio of quark shadowing to gluon
shadowing saturate?

\begin{figure}
\epsfxsize=9.0cm
\epsffile{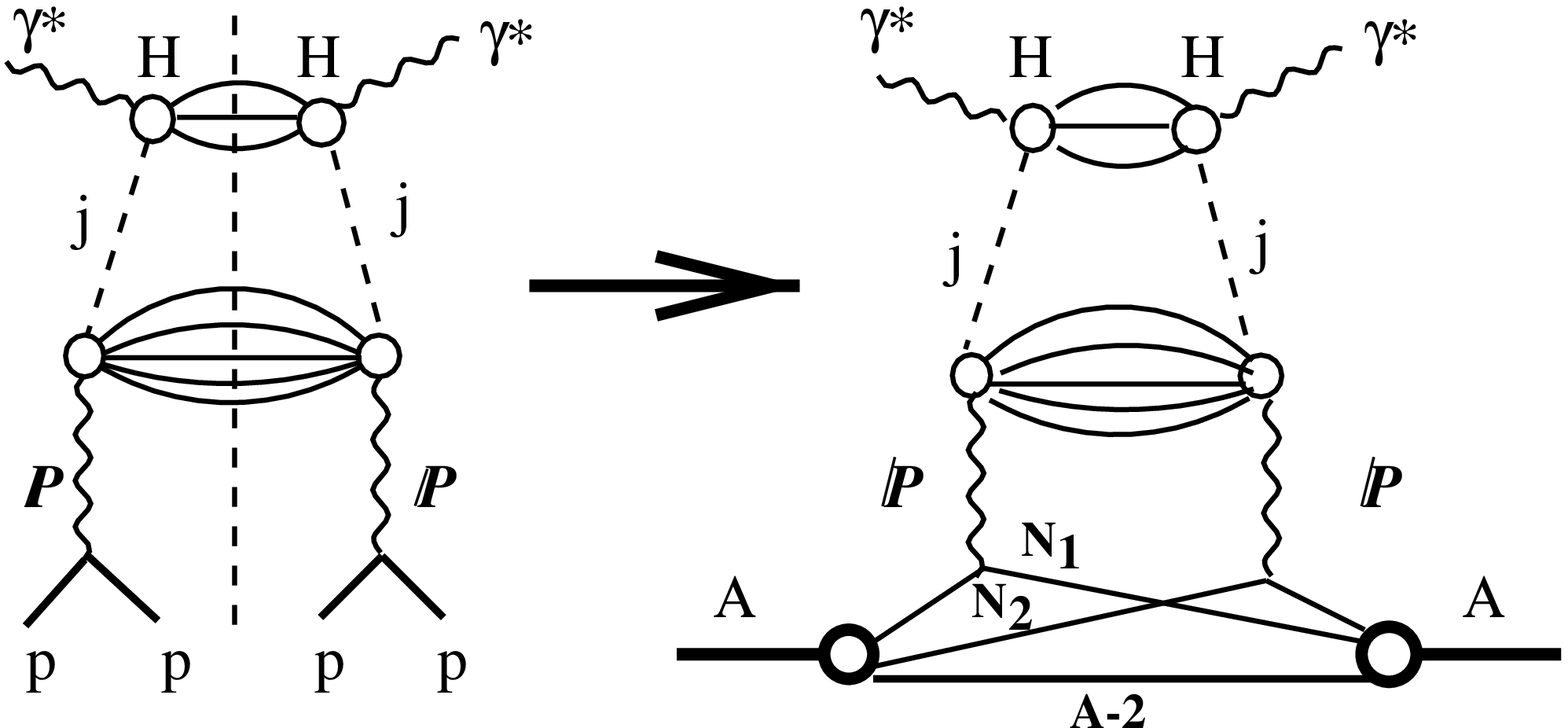}
\caption{Diagrams demonstrating the relation between gluon induced 
hard diffraction on protons and the leading twist contribution to nuclear 
shadowing in DIS. Figure from Ref.~\cite{FGS}.}
\label{fig:shad}
\end{figure}

Finally, it is well known that there is a close relation between
shadowing and diffraction. See Fig.~\ref{fig:shad}.  
Whether this relation persists
at  high parton densities is not known. In an interesting recent
exercise, it has been shown that diffractive {\it nucleon} data at
HERA could be used to predict the shadowing of quark distributions
observed by NMC~\cite{Kaidalov,FS}.  
Significant deviations from the simple relation
between shadowing and diffraction, may again suggest the presence of
strong non--linearities. At eRHIC the validity of this relation can be
explored directly--different nuclear targets are available, and the diffractive structure function may also be measured
independently.

\vskip 0.05in

\noindent\underline{Semi--inclusive signatures of the CGC}
\vskip 0.05in
The discussion of the relation between shadowing and diffraction provides 
a smooth segue into the topic of semi--inclusive signatures of the 
Colored Glass Condensate. While the 
novel physics of saturation and the formation 
of a CGC should be visible in inclusive quantities, their most dramatic 
manifestation will be in semi--inclusive measurements.

The most striking of these semi--inclusive measurements is hard
diffraction. Hard diffraction is the phenomenon wherein the virtual photon 
emitted
by the electron fragments into a final state $X$, with an invariant
mass $M_X^2\gg \Lambda_{QCD}^2$, while the proton emerges unscathed in
the interaction. A large rapidity gap--a region in rapidity essentially
devoid of particles--is produced between the fragmentation region of
the electron and that of the proton.  In pQCD, the probability of a
gap is exponentially suppressed as a function of the gap size. At HERA
though, gaps of several units in rapidity are unsuppressed; one finds
that roughly 10\% of the cross--section corresponds to hard
diffractive events with invariant masses $M_X > 3$ GeV. The remarkable
nature of this result is transparent in the proton rest frame; a $50$ TeV
electron slams into the proton and, 10\% of the time, the proton is
unaffected, even though the interaction causes the virtual photon to
fragment into a hard final state.

\begin{figure}
\epsfxsize=9.0cm
\epsffile{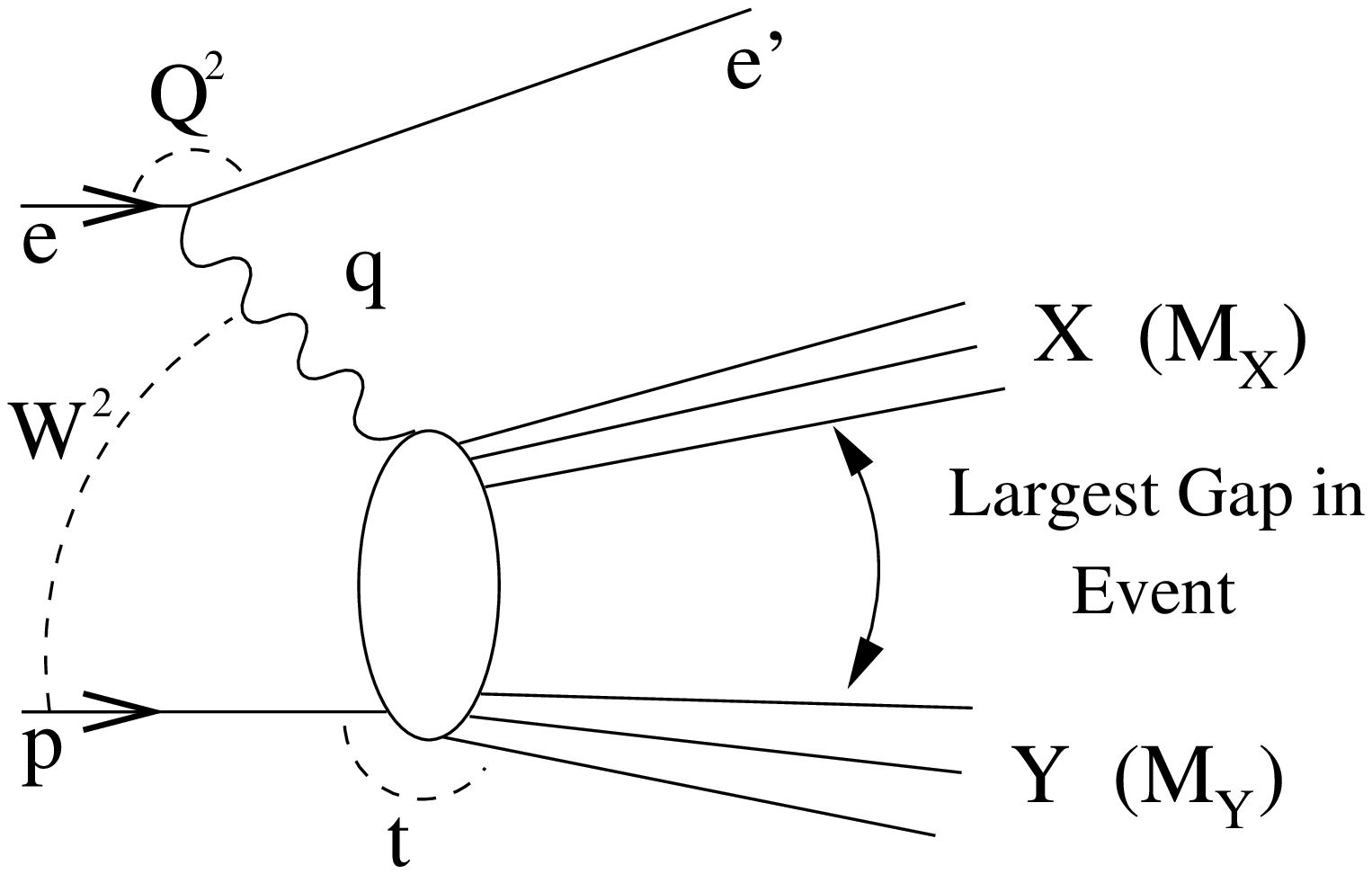}
\caption{The diagram of a process with a rapidity gap between the systems 
X and Y. The projectile nucleus is denoted here as p. Figure from Ref.~\cite{Arneodoetal}.}
\label{fig:madif}
\end{figure}

The interesting question in diffraction is to study the nature of the
color singlet object (the ``Pomeron'') within the proton that
interacts with the virtual photon since it addresses, in a novel
fashion, the central mystery of QCD--the nature of confining
interactions within hadrons. In hard diffraction, the mass of
the final state is large and one can reasonably ask questions about the
quark and gluon content of the Pomeron. A diffractive structure function 
$F_{2,A}^{D(4)}$ can be defined~\cite{BereraSoper,Veneziano,Collins}, in a fashion analogous to $F_2$, as
\be
{d^4 \sigma_{eA\rightarrow eXA}\over {dx_{Bj} dQ^2 d\xp dt}} & &=
A\cdot {4\pi \alpha_{em}^2\over x Q^4}\,\nonumber \\
& & \left\{ 1-y + 
{y^2\over 2[1+ R_A^{D(4)}(\beta,Q^2,\xp,t)]}\right\}\, F_{2,A}^{D(4)}
(\beta,Q^2,\xp,t) \, ,
\ee
where, $y=Q^2/s x_{Bj}$, and analogously to $F_2$, one has  
$R_A^{D(4)}=F_L^{D(4)}/F_T^{D(4)}$. Also,   
\be
Q^2= -q^2 > 0\,\,;\,\, x_{Bj} = {Q^2\over 2P\cdot q}\,\,;\,\,
\xp = {q\cdot (P-P^\prime)\over q\cdot P}\,\,;\,\, t = (P-P^\prime)^2 \, ,
\ee
and $\beta = x_{Bj}/\xp$. Here $P$ is the initial nuclear momentum, and 
$P^\prime$ is the net momentum of the fragments $Y$ in the proton 
fragmentation region. Similarly, $M_X$ is the net momentum of the 
fragments $X$ in the electron fragmentation region. An illustration of 
the hard diffractive event is shown in Fig.~\ref{fig:madif}. 
Unlike $F_2$ however, 
$F_2^{D(4)}$ is not truly universal--it cannot be applied, for instance, 
to predict diffractive cross--sections in $p$--$A$ scattering; it can 
be applied only in other lepton--nucleus scattering studies~\cite{Collins}.

It is more convenient in practice to measure the structure 
function $F_{2,A}^{D(3)}=\int F_{2,A}^{D(4)} dt$, where 
$|t_{min}|<|t|<|t_{max}|$, where $|t_{min}|$ is the minimal momentum 
transfer to the nucleus, and $|t_{max}|$ is the maximal momentum transfer 
to the nucleus that still 
ensures that the particles in the nuclear fragmentation region $Y$ are 
undetected. An interesting quantity to measure is the ratio 
\be
R_{A1,A2}(\beta,Q^2,\xp) = {F_{2,A1}^{D(3)}(\beta,Q^2,\xp)\over 
F_{2,A2}^{D(3)}(\beta,Q^2,\xp)} \, .
\ee
For eA at HERA, it was argued that this ratio could be measured with high 
systematic and statistical accuracy~\cite{Arneodoetal}--the situation for eRHIC should be 
at least comparable, if not better. If $R_{A1,A2} =1$, one can conclude 
that the structure of the Pomeron is universal, and one has an 
$A$--independent Pomeron flux. If $R_{A1,A2} = f(A1,A2)$, then albeit 
a universal Pomeron structure, the flux is A--dependent. Finally, if 
Pomeron structure is $A$--dependent, some models argue that 
$R_{A1,A2} = F_{2,A1}/F_{2,A2}$.

We discussed previously that the ratio of 
$R_D=\sigma_{\rm{diffractive}}/\sigma_{\rm{total}}$ at HERA is 
$\sim 10$\% for $M_X > 3$ GeV. The systematics of 
hard diffraction at HERA can be understood in saturation models~\cite{GBW}. 
For eA collisions at eRHIC energies, saturation
models predict that the ratio $R_D^A$ can be much higher--on the order 
of $30$\% for the largest nuclei~\cite{LevMaor,FS1}. 
The appearance of a large rapidity 
gap in $30$\% of all eA scattering events would be a striking confirmation 
of the saturation picture. Much theoretical and experimental work 
needs to be done to flesh out the details of predictions for hard 
diffractive events at eRHIC. Recent estimates suggest, for instance, that 
$F_L^D/F_T^D$, like $F_L/F_T$ may have a peak as a function of $Q^2$, whose 
position, likewise, increases with decreasing $x_{Bj}$ and increasing 
$A$~\cite{BNLTA}.

An important semi--inclusive observable in eA DIS at high 
energies is coherent (or diffractive) and inclusive vector meson production.  
As discussed by Brodsky et al.~\cite{Brodsky} (see also Ref.~\cite{KNNZ}), 
the forward vector meson diffractive leptoproduction cross--section off 
nuclei 
\be
{\frac{d\sigma}{dt}}|_{t=0}(\gamma^{*}A\rightarrow VA) \propto \alpha_S^2(Q^2)\,\left[G_A(x,Q^2)\right]^2 
\, ,
\ee
for large $Q^2$. Here $V$ denotes the vector meson. This quantity is clearly 
very sensitive to the gluon structure function. The ratio of this quantity 
in nuclei to that in nucleons is therefore (like the ratio of the 
longitudinal structure function) a probe of gluon shadowing.

In the color dipole picture, the amplitude for diffractive leptoproduction 
can be written as a convolution of the $q{\bar q}$ component of the 
$\gamma^{*}$ wavefunction times the $q{\bar q}$--nucleus cross--section 
times the vector meson wavefunction. In the saturation picture, a 
semi--hard scale is introduced via the $q{\bar q}$--A cross-section--whether 
this scale is larger or smaller than the scale associated with the 
size of the vector meson strongly affects the energy and $Q^2$ dependence 
of the vector meson cross-sections at small $x_{Bj}$. 
Recently, Caldwell and Soares~\cite{CS}
have studied vector meson production at HERA in the Golec-Biernat--Wusthoff 
model of saturation~\cite{GBW}. They find that the model provides a 
good description of the cross-section for photo-and electro-production 
of $J/\Psi$ in a wide $Q^2$ and energy range. For $\rho$ meson production, 
the change in the energy dependence as a function of $Q^2$ is well described 
by the model but the normalization of the cross-section is not. One 
possible explanation of this discrepancy is the lack of knowledge about 
the $\rho$ wavefunction.

We should  mention here that it is very important to measure 
inclusive and diffractive open charm  and jets since they provide 
useful and independent measures of the gluon distribution (and 
gluon shadowing) at small 
$x_{Bj}$. These will complement information on the gluon structure functions 
obtained from the $\ln(Q^2)$ derivative of $F_2$, from $F_L$, and 
from diffractive leptoproduction of vector mesons.

We may conclude from the above that 
it is in the measurement of semi--inclusive observables that a future 
collider environment has a marked superiority over previous fixed 
target electron--nucleus experiments. Rapidity gaps in eA collisions 
will be measured for the first time. Coherent and incoherent 
vector meson production can be studied in great detail in a wide 
kinematic range with much greater accuracy than previously.

At small $x_{Bj}$, the high parton densities produce large color fluctuations 
which are subsequently reflected in large multiplicity fluctuations. One 
expects for instance the following phenomena: a) a broader rapidity 
distribution in larger nuclei relative to lighter nuclei and protons, 
b) Rapidity correlations over several units of rapidity--an anomalous 
multiplicity in one rapidity interval in an event would be accompanied by 
an anomalous multiplicity in rapidity intervals several units 
away~\cite{AGK,KLM} 
and c) a correlation between the central multiplicity with the multiplicity 
of neutrons in a forward neutron detector~\cite{FS}.

\subsection{Precision measurements of nuclear observables at small $x_{Bj}$}

In much of our discussion, we have focused on the potential of 
an eA collider to discover a novel state of saturated gluonic matter--
the colored glass condensate. This search, if successful, could revolutionize 
our understanding of QCD at high energies by providing answers to 
questions about the nature of confinement in high energy scattering, the 
origins of multi--particle production, the asymptotic behavior of 
cross--sections, etc.

However, even in the absence of the promise of radically new physics, 
there is a compelling  case to be made for 
an eA collider. The gluon distribution in a nucleus is ill--understood. 
Understanding its behavior as a function of $x_{Bj}$ and $Q^2$, and 
of its shadowing is of intrinsic interest. At eRHIC, as discussed above, 
measurements of $F_L$, and of semi--inclusive quantities promise that 
the gluon distribution in a nucleus could be independently extracted 
with high precision. The nuclear gluon distribution extracted with 
an eA collider can be compared with the distribution extracted from AA 
and pA collisions. 

The $A$ dependence (as a function of $x_{Bj}$ and $Q^2$) of vector 
meson production at small $x_{Bj}$ ]
is also of intrinsic interest, as well of use in interpretations of 
pA and AA collisions--a particular example being that of $J/\psi$ 
suppression. 

Hard diffraction off nuclei has not been previously measured. At eRHIC, 
nuclear diffractive structure functions can be measured for the 
first time. The relation of these to $F_2$ will, as discussed previously, 
provide new insight into the relation between diffraction and shadowing.

At intermediate $x$'s, an eA collider provides a laboratory to 
study the propagation of fast partons through nuclear matter.
Color transparency and color opacity, which we have not 
discussed here, can be studied more extensively than 
previously~\cite{FKS}. Jet 
quenching, often cited as a signature of the quark gluon plasma 
in nuclear collisions~\cite{GPW}, can be investigated in the
cold nuclear environment of an eA collider~\cite{Genya} and compared 
to results from AA collisions.

\subsection{Connections to pA and AA physics}

We will very briefly discuss here the relation of the physics of an 
eA collider to pA and AA physics at current and future collider facilities.

\vskip 0.05in
\noindent\underline{Relation of eA to pA}

In pA scattering at RHIC~\footnote{For a recent discussion, see the
proceedings of the pA workshop at BNL, Oct. 28th--29th,
Eds. S. Aronson and J. C. Peng, at the website 
www.bnl.gov/rhic/townmeeting/agenda$_b$.htm}, one also has the
opportunity to study the gluon distribution in nuclei. Gluon fusion to
jets, vector mesons, open charm and beauty can be measured. Hard and
soft diffraction--the size and distribution of energy gaps with energy
and nuclear size can also be studied. Scaling violations in Drell--Yan
scattering can be measured for the first time.

Some of the differences between pA and eA are as follows. In pA scattering, 
for instance in the signature Drell--Yan process, it is very hard to 
reliably extract distributions in the region below the $\Psi^\prime$ tail--
namely, one requires $Q^2 > 16$ GeV$^2$. In the $x$ region of interest, 
one expects saturation effects to be important at lower $Q^2$ of 
$1$--$10$ GeV$^2$. For $Q^2=16$ GeV$^2$, one might have to go to significantly 
smaller $x$'s to see large saturation effects. Secondly, the survival 
probability of large rapidity gaps is smaller in pA relative to eA. This is 
because the gap is destroyed due to secondary interactions between 
``spectator'' 
partons in the proton and the ``Pomeron'' from the nucleus. This does 
not occur in eA scattering because of course there are no spectator 
partons in the electron. Thus one expects that diffractive vector meson and 
jet production in pA should be qualitatively different than what one 
will see in eA.

\vskip 0.05in
\noindent\underline{Relation of eA to AA}

A large variety of models combining hard and soft physics are used 
to study nuclear collisions at RHIC and LHC energies~\cite{Pajares}.
Many of these model predictions depend sensitively on the nuclear gluon 
density--for a recent parametrization of nuclear gluon densities, see
Ref.~\cite{EKS98}. Data from an eA collider will be essential in 
further refining these parametrizations.

In the classical approach discussed previously, the relation between
the parton distributions in the nuclear wavefunction and the
multiplicity of produced gluons simplifies--the initial multiplicity
of produced gluons is given in terms of the saturation scale $Q_s$ by
the simple relation~\cite{Mueller2}
\be
\frac{1}{\pi R^2}\, \frac{dN}{d\eta} = c_N \, \frac{N_c^2-1}{N_c}\,\frac{1}
{4\pi^2\alpha_S}\, Q_s^2 \, .
\ee
The coefficient $c_N$ can be estimated numerically in classical
lattice simulations of nuclear collisions~\cite{AlexRaju1} and is
determined to be $c_N\sim 1.3$.  A similar analysis is used to
determine the initial energy of the produced
glue~\cite{AlexRaju2}. This distribution is only the initial parton
distribution--the subsequent possible evolution to a quark gluon
plasma~\cite{thermal} is controlled again only by the scale $Q_s$.

It is therefore conceivable that high energy heavy ion collisions, 
despite their complexity, may provide insight into the parton distributions 
in the nuclear wavefunction. An eA collider will confirm and deepen our 
understanding of what we may learn from heavy ion collisions.

\section{Summary}

In this talk, we discussed the physics case for an eA collider. We 
emphasized the novel physics that might be studied at small $x$. 
The interesting physics at intermediate $x$'s has been discussed 
elsewhere~\cite{Arneodoetal}. 

Plans for an electron--ion collider 
include, as a major part of the program, the possibility of doing 
polarized electron--polarized proton/light ion scattering. A 
discussion 
of the combined case for high energy electron nucleus  and 
polarized electron--polarized proton scattering will be published 
separately~\cite{AbhayRaju}.

\section{Acknowledgement}

I would like to thank the organizers of the EPIC meeting at MIT for
inviting me to present the physics case for eRHIC. Many thanks to
Abhay Deshpande, Witek Krasny, Yuri Kovchegov, Genya Levin, Larry
McLerran, and Mark Strikman for discussions, and for kindly allowing me to
use figures from their papers. In particular, I would like to thank
Mark Strikman for his detailed comments on the manuscript.  I would
also like to thank all the participants of the eA meeting at BNL in
June/July 2000 for the stimulating discussions that have informed this
report. This work was supported at BNL by the Department of Energy
under DOE contract number DOE-FG02-93-ER-40764, by RIKEN--BNL, 
and by an LDRD grant from Brookhaven Science Associates.


\begin{thebibliography}{99}

\bibitem{proceed1}Proceedings of the 2nd eRHIC workshop, Yale, April 6th--8th, 
2000, BNL--52592.

\bibitem{proceed2}Proceedings of the BNL summer meeting, June 26th--July 14th, 
2000, BNL--52606

\bibitem{Arneodoetal}M. Arneodo et al., in Proceedings of {\it Future 
Physics at HERA}, DESY, September 25th--26th, 1995, hep-ph/9610423. 

\bibitem{largex}proceedings of the EPIC meeting, MIT, Boston, September 
14th-16th, 2000.

\bibitem{Pramana}R.~Venugopalan,
Pramana {\bf 55}, 73 (2000)
[hep-ph/0005096].

\bibitem{LowNussinov}F.~E.~Low,
Phys.\ Rev.\ D {\bf 12}, 163 (1975);
S.~Nussinov,
Phys.\ Rev.\ D {\bf 14}, 246 (1976).

\bibitem{DL}
A.~Donnachie and P.~V.~Landshoff,
Phys.\ Lett.\ {\bf B296}, 227 (1992)
[hep-ph/9209205].

\bibitem{Arneodo}M.~Arneodo,
Phys.\ Rept.\ {\bf 240} (1994) 301.

\bibitem{Lipatov}E.~A.~Kuraev, L.~N.~Lipatov and V.~S.~Fadin,
Sov.\ Phys.\ JETP{\bf 45}, 199 (1977);
I.~I.~Balitsky and L.~N.~Lipatov,
Sov.\ J.\ Nucl.\ Phys.\ {\bf 28}, 822 (1978).

\bibitem{FadLip}V.~S.~Fadin and L.~N.~Lipatov,
Phys.\ Lett.\ {\bf B429}, 127 (1998)
[hep-ph/9802290].

\bibitem{CiaCam}G.~Camici and M.~Ciafaloni,
Phys.\ Lett.\ {\bf B412}, 396 (1997)
[hep-ph/9707390].

\bibitem{Salam}G.~P.~Salam,
Acta Phys.\ Polon.\ {\bf B30}, 3679 (1999)
[hep-ph/9910492].

\bibitem{GLR}L.~V.~Gribov, E.~M.~Levin and M.~G.~Ryskin,
Phys.\ Rept.\ {\bf 100}, 1 (1983).

\bibitem{MQ}A.~H.~Mueller and J.~Qiu,
Nucl.\ Phys.\ {\bf B268}, 427 (1986).

\bibitem{MV}L.~McLerran and R.~Venugopalan,
Phys.\ Rev.\ D {\bf 49}, 2233 (1994); {\it ibid.}, 3352 (1994); 
Phys.\ Rev.\ D {\bf 50}, 2225 (1994); 
Phys.\ Rev.\ {\bf D 59}, 094002 (1999).


\bibitem{DeRujula}A.~De Rujula, S.~L.~Glashow, H.~D.~Politzer, S.~B.~Treiman, F.~Wilczek and A.~Zee,
Phys.\ Rev.\ D {\bf 10}, 1649 (1974); D.~J.~Gross,
Phys.\ Rev.\ Lett.\ {\bf 32}, 1071 (1974).

\bibitem{BallForte}R.~D.~Ball and S.~Forte,
Phys.\ Lett.\ {\bf B358}, 365 (1995)
[hep-ph/9506233].

\bibitem{PDF}H.~L.~Lai {\it et al.}  [CTEQ Collaboration],
Eur.\ Phys.\ J.\ {\bf C12}, 375 (2000)
[hep-ph/9903282]; A.~D.~Martin, R.~G.~Roberts, W.~J.~Stirling and R.~S.~Thorne,
Nucl.\ Phys.\ Proc.\ Suppl.\ {\bf 79}, 105 (1999)
[hep-ph/9906231]; M.~Gluck, E.~Reya and A.~Vogt,
Eur.\ Phys.\ J.\ {\bf C5}, 461 (1998)
[hep-ph/9806404].


\bibitem{ManSaaWei}L.~Mankiewicz, A.~Saalfeld and T.~Weigl,
hep-ph/9706330.

\bibitem{AltBalFort}G.~Altarelli, R.~D.~Ball and S.~Forte,
hep-ph/0011270.

\bibitem{ZEUSa}J.~Breitweg {\it et al.}  [ZEUS Collaboration],
Eur.\ Phys.\ J.\ {\bf C6}, 43 (1999)
[hep-ex/9807010].

\bibitem{H1a}C.~Adloff {\it et al.}  [H1 Collaboration],
Z.\ Phys.\ {\bf C76}, 613 (1997)
[hep-ex/9708016].

\bibitem{McDermott}M.~F.~McDermott,
hep-ph/0008260.

\bibitem{TelAviv}E.~Gotsman, E.~Levin, M.~Lublinsky, U.~Maor, 
E.~Naftali and K.~Tuchin,
hep-ph/0010198.

\bibitem{FDS}L.~Frankfurt, M.~McDermott and M.~Strikman,
hep-ph/0009086.

\bibitem{GBW}
K.~Golec-Biernat and M.~Wusthoff,
Phys.\ Rev.\ D {\bf 59}, 014017 (1999)
[hep-ph/9807513]; 
{\it ibid.}, {\bf 60}, 114023 (1999)
[hep-ph/9903358].

\bibitem{GLM}
E.~Gotsman, E.~Levin, U.~Maor and E.~Naftali,
Nucl.\ Phys.\ {\bf B539}, 535 (1999)
[hep-ph/9808257].

\bibitem{FGS}L.~Frankfurt, V.~Guzey and M.~Strikman,
J.\ Phys.\ G{\bf G27}, R23 (2001)
[hep-ph/0010248].


\bibitem{BKS}A.~M.~Stasto, K.~Golec-Biernat and J.~Kwiecinski,
hep-ph/0007192.

\bibitem{Mueller1}A.~H.~Mueller, 
Lectures given at {\it 
International Summer School on Particle Production Spanning MeV
and TeV Energies}, Nijmegen, Netherlands, 8-20 Aug 1999, 
hep-ph/9911289.

\bibitem{Larry}E.~Iancu, A.~Leonidov and L.~McLerran,
hep-ph/0011241.

\bibitem{ParisiSourlas}G.~Parisi and N.~Sourlas,
Phys.\ Rev.\ Lett.\ {\bf 43}, 744 (1979).

\bibitem{JKMW}J.~Jalilian-Marian, A.~Kovner, L.~McLerran and H.~Weigert,
Phys.\ Rev.\ D {\bf 55}, 5414 (1997)
[hep-ph/9606337].

\bibitem{Kovchegov}Y.~V.~Kovchegov,
Phys.\ Rev.\ D {\bf 54}, 5463 (1996)
[hep-ph/9605446].

\bibitem{KovMuell}
Y.~V.~Kovchegov and A.~H.~Mueller,
Nucl.\ Phys.\ {\bf B529}, 451 (1998)
[hep-ph/9802440].

\bibitem{JKW}J. Jalilian--Marian, A. Kovner and H. Weigert, 
{\em Phys. Rev.} {\bf D59} 014015 (1999).

\bibitem{RG}I. Balitsky, {\em Phys. Rev.} {\bf D60} 014020 (1999); 
{\em Phys. Rev. Lett.} {\bf 81} 2024 (1998); Y.~V.~Kovchegov, 
{\em Phys. Rev.}  {\bf D61}, 074018 (2000); E.~Levin and K.~Tuchin, 
hep-ph/9908317;  E.~Iancu, A.~Leonidov and L.~McLerran,
hep-ph/0011241.

\bibitem{DGLAP}V.~N.~Gribov and L.~N.~Lipatov, Sov.\ J.\ Nucl.\ Phys.\ 
{\bf 15}, 78 (1972); G.~Altarelli and G.~Parisi,
Nucl.\ Phys.\ {\bf B126}, 298 (1977); Yu.~L.~Dokshitzer, Sov.\ Phys.\ JETP.\
{\bf 73}, 1216 (1977).

\bibitem{StrikZhalov}M.~Strikman, M.~G.~Tverskoii and M.~B.~Zhalov,
Phys.\ Lett.\ {\bf B459}, 37 (1999)
[nucl-th/9806099].

\bibitem{Krasny}W. Krasny, contribution in Ref. 1.

\bibitem{Wilczeketal}A. Zee, F. Wilczek, and S. B. Treiman, Phys.\ Rev.\ D
{\bf 10}, 2881 (1974); W. Bardeen, A. J. Buras, D. W. Duke, and T. Muta, 
Phys.\ Rev.\ D {\bf 18}, 3998 (1978).

\bibitem{Bartels}J.~Bartels, K.~Golec-Biernat and K.~Peters,
Eur.\ Phys.\ J.\ {\bf C17}, 121 (2000)
[hep-ph/0003042].

\bibitem{BNLTA}E.~Gotsman, E.~Levin, U.~Maor, L.~McLerran and K.~Tuchin,
hep-ph/0008280; hep-ph/0007258.


\bibitem{Brodsky}S.~J.~Brodsky, L.~Frankfurt, J.~F.~Gunion, A.~H.~Mueller and M.~Strikman,
Phys.\ Rev.\ {\bf D 50}, 3134 (1994)
[hep-ph/9402283].

\bibitem{KNNZ}B.~Z.~Kopeliovich, J.~Nemchick, 
N.~N.~Nikolaev and B.~G.~Zakharov,
Phys.\ Lett.\ {\bf B324}, 469 (1994)
[hep-ph/9311237].


\bibitem{JamalWang}
J.~Jalilian-Marian and X.~Wang,
Phys.\ Rev.\ D {\bf 60}, 054016 (1999)
[hep-ph/9902411].


\bibitem{Kaidalov}A.~Capella, A.~Kaidalov, C.~Merino, D.~Pertermann and 
J.~Tran Thanh Van,
Eur.\ Phys.\ J.\ {\bf C5}, 111 (1998)
[hep-ph/9707466].

\bibitem{BereraSoper}A.~Berera and D.~E.~Soper,
Phys.\ Rev.\ D {\bf 53}, 6162 (1996)
[hep-ph/9509239].

\bibitem{Veneziano}L.~Trentadue and G.~Veneziano,
Phys.\ Lett.\ {\bf B323}, 201 (1994).


\bibitem{Collins}J.~C.~Collins,
Phys.\ Rev.\ D {\bf 57}, 3051 (1998)
[hep-ph/9709499].


\bibitem{LevMaor}E.~Levin and U.~Maor,
hep-ph/0009217.

\bibitem{FS1}L. Frankfurt and M. Strikman, {\em Phys. Lett.} {\bf B382} 
(1996) 6.

\bibitem{CS}A.~C.~Caldwell and M.~S.~Soares,
hep-ph/0101085.

\bibitem{AGK}V.~A.~Abramovsky, V.~N.~Gribov and O.~V.~Kancheli,
Yad.\ Fiz.\ {\bf 18}, 595 (1973).

\bibitem{KLM}Y.~V.~Kovchegov, E.~Levin and L.~McLerran,
hep-ph/9912367.

\bibitem{FS}L.~Frankfurt and M.~Strikman,
Eur.\ Phys.\ J.\ {\bf A5}, 293 (1999)
[hep-ph/9812322].


\bibitem{FKS}L.~Frankfurt, V.~Guzey, W.~Koepf, M.~Sargsian and M.~Strikman,
hep-ph/9608492.


\bibitem{GPW}X.~Wang, M.~Gyulassy and M.~Plumer,
Phys.\ Rev.\ D {\bf 51}, 3436 (1995)
[hep-ph/9408344].

\bibitem{Genya}R.~Baier, Y.~L.~Dokshitzer, S.~Peigne and D.~Schiff,
Phys.\ Lett.\ {\bf B345}, 277 (1995)
[hep-ph/9411409];  M.~Luo, J.~W.~Qiu and G.~Sterman,
Phys.\ Rev.\ D {\bf 50}, 1951 (1994).  E.~Levin,
Phys.\ Lett.\ {\bf B380}, 399 (1996)
[hep-ph/9508414]. 

\bibitem{Pajares}N.~Armesto and C.~Pajares,
Int.\ J.\ Mod.\ Phys.\ {\bf A15}, 2019 (2000)
[hep-ph/0002163].

\bibitem{EKS98}K.~J.~Eskola, V.~J.~Kolhinen and C.~A.~Salgado,
Eur.\ Phys.\ J.\ {\bf C9}, 61 (1999)
[hep-ph/9807297].

\bibitem{Mueller2}A.~H.~Mueller,
Nucl.\ Phys.\ {\bf B572}, 227 (2000) [hep-ph/9906322].

\bibitem{AlexRaju1}A.~Krasnitz and R.~Venugopalan,
hep-ph/0007108, {\it Phys. Rev. Lett.} in press.

\bibitem{AlexRaju2}A.~Krasnitz and R.~Venugopalan,
Phys.\ Rev.\ Lett.\ {\bf 84}, 4309 (2000) [hep-ph/9909203].

\bibitem{thermal}A.~Dumitru and M.~Gyulassy,
Phys.\ Lett.\ {\bf B494}, 215 (2000)
[hep-ph/0006257]; J.~Bjoraker and R.~Venugopalan,
Phys.\ Rev.\ {\bf C 63}, 024609 (2001)
[hep-ph/0008294]; R.~Baier, A.~H.~Mueller, D.~Schiff and D.~T.~Son,
hep-ph/0009237.

\bibitem{AbhayRaju}A.~Deshpande and R.~Venugopalan, in preparation.


\end{thebibliography}
\end{document}